\documentclass[pra]{revtex4}

\usepackage{graphicx}
\usepackage{dcolumn}
\usepackage{bm}

\begin{document}


\title{Intrinsic decoherence effects on tripartite GHZ state generation 
using a trapped ion coupled to an optical cavity}

\author{S. Shelly Sharma}
\email[shelly@uel.br]{shelly@uel.br}
\thanks{}
\affiliation{Depto. de Fisica, Universidade Estadual de Londrina,
Londrina 86040-370, PR Brazil }

\author{N. K. Sharma}
\email[nsharma@uel.br]{Your e-mail address}
\thanks{}
\affiliation{Depto. de Matematica, Universidade Estadual de Londrina,
Londrina 86040-370 PR, Brazil }

\date{\today}

\begin{abstract}
We analyse the effects of intrinsic decoherence on the probability of
generating a tripartite GHZ state using a cool trapped ion coupled to 
a single mode of the cavity field and interacting with a resonant laser field. 
Milburn equation is solved for this tripartite system to obtain the time evolution 
of density matrix as a function of cavity-ion coupling, laser-ion coupling, and the 
size of unitary time step relative to the time scale determined by system parameters. 
Starting with the system prepared initially in a separable state, density matrix is 
used to calculate the probability of tripartite GHZ state generation using coupling 
strengths reported in a recent experiment.

\end{abstract}

\maketitle

Interest in multipartite maximally entangled states is motivated by possible
use in quantum communication \cite{eker91}, quantum dense coding \cite
{benn92} and quantum teleportation \cite{benn93}. Advances in experimental
techniques of trapping, cooling, preparation and manipulation of ionic
quantum states\cite{monr95,wine98,roos99}, through interaction with external
lasers, have made the generation of maximally entangled states a
possibility. Coupling of trapped ion to quantized field inside an optical
cavity has also been achieved \cite{mund03}. With the ion trap placed inside
a high finesse cavity, we have at hand a tripartite system with additional
control mechanism offered by ion cavity coupling. This is the basic unit of
a more complex multiple function system in which quantum states of several
trapped ions placed coupled to an optical cavity may be manipulated in a
controlled manner. In a letter\cite{shar103}, we proposed a scheme to
generate three qubit maximally entangled GHZ state, using trapped ion
interacting with a resonant external laser and sideband tuned single mode of
a cavity field. Experimental implementation of such a proposal is
constrained by the effects of quantum decoherence sources. Suppression of
quantum coherence occurs due to random and unknown perturbations of system
Hamiltonian. In a GHZ state generation experiment, the average effect of
decoherence sources on probability of finding the tripartite system in GHZ
state is of interest. In this paper, we investigate the effects of intrinsic
decoherence on the composite system dynamics using Milburn model\cite{milb91}%
. Milburn model of intrinsic decoherence is based on the assumption that a
closed quantum system evolves in a stochastic sequence of identical unitary
transformations. The size of the unitary step along with the energy spectrum
of the quantum system, determine the decoherence rate. The model applied by
Milburn to free evolution of a closed quantum system, has since been used to
study decoherence effects in JCM model with a single quantized field mode 
\cite{cess93,plen97}, decoherence in multiphoton JCM model\cite{kuan95},
decoherence of open systems\cite{buze98}, decoherence of a two-level
particle coupled to photon field\cite{zida02} and decoherence of quantum
gate \cite{kimm02}. We solve the Milburn equation exactly for a cool trapped
ion coupled to a single mode of the cavity field and interacting with a
resonant laser field. The evolution dynamics of tripartite system is a
function of laser-ion coupling, cavity ion coupling, and size of unitary
step relative to the time scale determined by system parameters. The system\
is prepared initially in a separable state having ionic center of mass and
cavity mode in respective number states, while the ion is in ground state.
From the density matrix containing intrinsic decoherence effects, we
calculate the probability of tripartite GHZ state generation using coupling
strengths reported in recent experiments\cite{mund03}.

Cold ions in a linear trap \cite{monr95,wine98,roos99} offer a promising
physical system for quantum information processing, as each ion allows two
qubit state manipulation. With an ion trap placed inside a high finesse
optical cavity, we obtain a three qubit system with additional control
mechanism offered by quantized cavity field. The interaction picture
Hamiltonian for this system is given in a detailed form in Eq. (4) of ref. 
\cite{shar203}. On the time scale involving only a few cycles of unitary
time evolution, the effective dimensions of three subsystems namely ionic
internal state ($A$), state of center of mass motion ($B$), and cavity field
state ($C$) depend not only on the ion type but also on the relevant part of
the ion-field interaction. Presently, we consider the resonant interaction
of a two level cold trapped ion having transition frequency $\omega _{0},$
with an external laser of frequency $\omega _{L}=\omega _{0},$ while the
cavity field is tuned to red sideband of ionic vibrational motion, $\omega
_{0}-\omega _{c}=\nu $. Here $\nu $ is the frequency of one dimensional
harmonic oscillator trapping potential. The Lamb-Dicke (LD) parameters
relative to the laser field and the cavity field are denoted by $\eta _{L}$
and $\eta _{c}$ respectively. Working in the Lamb-Dicke regime, $\eta
_{L}\ll 1$ and $\eta _{c}\ll 1$, the dynamics of the trapped two-level ion
is governed by interaction Hamiltonian, 
\begin{equation}
\hat{H_{I}}=\hbar \Omega \lbrack \sigma _{+}+\sigma _{-}]+\hbar g{\eta _{c}}%
\left[ \sigma _{+}\hat{b}\hat{a}+\sigma _{-}\hat{b}^{\dagger }\hat{a}%
^{\dagger }\right] ,  \label{2.1}
\end{equation}
where $\hat{a}^{\dagger }(\hat{a})$ and $\hat{b}^{\dagger }(\hat{b})$ are
creation (annihilation) operators for vibrational phonon and cavity field
photon respectively. The ion-phonon and ion-cavity coupling constants are $%
\Omega $ and $g$, whereas $\sigma _{k}(k=z,+,-)$ are the Pauli operators
qualifying the internal state of the ion.

To obtain time evolution of the system, we work in the basis $\left|
g,m,n\right\rangle ,$ $\left| e,m,n\right\rangle ,$ $\left|
g,m-1,n-1\right\rangle ,$ and $\left| e,m-1,n-1\right\rangle .$ Here $%
m,n=0,1,..,\infty $ denote the state of ionic vibrational motion and
quantized cavity field, respectively. The matrix representation of
interaction operator $\hat{H_{I}}$ in the chosen basis, to the lowest order
in $\eta _{L}$ and $\eta _{c},$ is 
\begin{equation}
H_{I}=\left[ 
\begin{array}{cccc}
0 & \hbar \Omega & 0 & \hbar g\eta _{c}\sqrt{mn} \\ 
\hbar \Omega & 0 & 0 & 0 \\ 
0 & 0 & 0 & \hbar \Omega \\ 
\hbar g\eta _{c}\sqrt{mn} & 0 & \hbar \Omega & 0
\end{array}
\right] .  \label{2.2}
\end{equation}
The unitary transformation that diagonalizes $H_{I}$ is easily obtained and
yields the eigenvectors, $\widehat{H}_{I}\Phi _{p}=E_{p}\Phi _{p},$ ($p=1,4$%
). Milburn \cite{milb91} postulates that on sufficiently short time steps,
the quantum system evolves stochastically and the change in density operator
in a time interval $(t,t+\tau )$ is given by

\begin{equation}
\widehat{\rho }(t+\tau )=\gamma \exp \left( -\frac{i}{\hbar }\theta (\tau )%
\widehat{H}_{I}\right) \widehat{\rho }(t)\exp \left( \frac{i}{\hbar }\theta
(\tau )\widehat{H}_{I}\right) ,  \label{2.4}
\end{equation}
$\theta (\tau )$ being some function of $\tau .$ In standard quantum
mechanics $\theta (\tau )=\tau .$ In Milburn model $\lim\limits_{\tau
\rightarrow 0}\theta (\tau )=1/\gamma $, a constant minimum time step with
mean frequency $\gamma $. The evolution of density operator $\widehat{\rho }%
(t)$ in Milburn model , is governed by the equation 
\begin{equation}
\frac{d}{dt}\widehat{\rho }(t)=\gamma \left[ \exp \left( -\frac{i}{\hbar
\gamma }\widehat{H}_{I}\right) \widehat{\rho }(t)\exp \left( \frac{i}{\hbar
\gamma }\widehat{H}_{I}\right) -\widehat{\rho }(t)\right] ,  \label{2.3}
\end{equation}
and the rate of change of density operator to first order in $\gamma ^{-1}$
as obtained from Eq. (\ref{2.3}) is 
\begin{equation}
\frac{d}{dt}\widehat{\rho }(t)=-\frac{i}{\hbar }\left[ \widehat{H}_{I},%
\widehat{\rho }(t)\right] -\frac{1}{2\hbar ^{2}\gamma }\left[ \widehat{H}%
_{I},\left[ \widehat{H}_{I},\widehat{\rho }(t)\right] \right] .  \label{2.5a}
\end{equation}

Transforming to the basis in which $\widehat{H}_{I}$ is diagonal and
integrating Eq. (\ref{2.5a}), we get 
\[
\left\langle \widehat{\rho }(t)\right\rangle =\sum_{p,q=1,4}\left\langle 
\widehat{\rho }(t)\right\rangle _{pq}\left| \Phi _{p}\right\rangle
\left\langle \Phi _{q}\right| 
\]
where 
\begin{equation}
\left\langle \widehat{\rho }(t)\right\rangle _{pq}=\left\langle \widehat{%
\rho }(0)\right\rangle _{pq}\exp \left( -\frac{i\left(
E_{p}-E_{q}\right) t }{\hbar } - \frac{(E_{p}-E_{q})^{2} t}{2\hbar^{2} \gamma }%
  \right) .  \label{2.8}
\end{equation}
\ The presence of\ factors of the type $\left( {E_{p}-E_{q}}/{\hbar }\right)
^{2}$ in the exponential indicates that energy spectrum of interaction
operator $\widehat{H}_{I}$ plays an important role in determining the
intrinsic decoherence rate. By solving Eq. (\ref{2.8}), for the initial
state $\widehat{\rho }(0)=\left| g,m-1,n-1\right\rangle \left\langle
g,m-1,n-1\right| $, we get
\begin{widetext} 
\begin{eqnarray}
\widehat{\rho }(t) &=&\frac{(A+B)^{2}}{2}\left[ \left| \Phi
_{1}\right\rangle \left\langle \Phi _{1}\right| +\left| \Phi
_{4}\right\rangle \left\langle \Phi _{4}\right| \right]  \nonumber \\
&&-\frac{(A+B)^{2}}{2}\left[ \exp \left( -\frac{2\left( \mu
_{mn}-a_{mn}\right) ^{2}t}{\gamma }-i2(\mu _{mn}-a_{mn})t\right) \left| \Phi
_{1}\right\rangle \left\langle \Phi _{4}\right| \right.  \nonumber \\
&&\left. +\exp \left( -\frac{2\left( \mu _{mn}-a_{mn}\right) ^{2}t}{\gamma }%
+i2(\mu _{mn}-a_{mn})t\right) \left| \Phi _{4}\right\rangle \left\langle
\Phi _{1}\right| \right]  \nonumber \\
&&+\frac{(A-B)^{2}}{2}\left[ \exp \left( -\frac{2\left( \mu
_{mn}+a_{mn}\right) ^{2}t}{\gamma }+i2(\mu _{mn}+a_{mn})t\right) \left| \Phi
_{2}\right\rangle \left\langle \Phi _{3}\right| \right.  \nonumber \\
&&\left. +\exp \left( -\frac{2\left( \mu _{mn}+a_{mn}\right) ^{2}t}{\gamma }%
-i2(\mu _{mn}+a_{mn})t\right) \left| \Phi _{3}\right\rangle \left\langle
\Phi _{2}\right| \right]  \nonumber \\
&&\frac{(A-B)^{2}}{2}\left( \left| \Phi _{2}\right\rangle \left\langle \Phi
_{2}\right| +\left| \Phi _{3}\right\rangle \left\langle \Phi _{3}\right|
\right)  \nonumber \\
&&+\frac{(A^{2}-B^{2})}{2}\left[ \exp \left( -\frac{2\mu _{mn}^{2}t}{\gamma }%
-i2\mu _{mn}t\right) \left( \left| \Phi _{1}\right\rangle \left\langle \Phi
_{2}\right| -\left| \Phi _{3}\right\rangle \left\langle \Phi _{4}\right|
\right) \right.  \nonumber \\
&&+\exp \left( -\frac{2\mu _{mn}^{2}t}{\gamma }+i2\mu _{mn}t\right) \left(
\left| \Phi _{2}\right\rangle \left\langle \Phi _{1}\right| -\left| \Phi
_{4}\right\rangle \left\langle \Phi _{3}\right| \right)  \nonumber \\
&&+\exp \left( -\frac{2a_{mn}^{2}t}{\gamma }+i2a_{mn}t\right) \left( \left|
\Phi _{1}\right\rangle \left\langle \Phi _{3}\right| -\left| \Phi
_{2}\right\rangle \left\langle \Phi _{4}\right| \right)  \nonumber \\
&&\left. +\exp \left( -\frac{2a_{mn}^{2}t}{\gamma }-i2a_{mn}t\right) \left(
\left| \Phi _{3}\right\rangle \left\langle \Phi _{1}\right| -\left| \Phi
_{4}\right\rangle \left\langle \Phi _{2}\right| \right) \right]  \label{2.9}
\end{eqnarray}
\end{widetext}
where $a_{mn}=\frac{1}{2}g\eta _{c}\sqrt{mn}$, $\mu _{mn}=\sqrt{%
a_{mn}^{2}+\Omega ^{2}}$, $A^{2}=(\mu _{mn}+\Omega )/4\mu _{mn}$, and $%
B^{2}=(\mu _{mn}-\Omega )/4\mu _{mn}$.

Next we consider a special case with the ion in its ground state occupying
the lowest energy trap state, while the cavity is prepared in vacuum state.
The density operator for the initial state is $\widehat{\rho }(0)=\left|
g,0,0\right\rangle \left\langle g,0,0\right| ,$($m=1,n=1).$ Without taking
into consideration the decoherence effects, the interaction time $t_{_{p}}$
needed to generate the maximally entangled tripartite GHZ state depends on
the the ratio $\alpha =\left( \mu _{11}/a_{11}\right) ,$ involving
parameters characterizing the ion- laser and ion-cavity interaction strengths%
$.$ For interaction time $t_{p}$ such that $a_{11}t_{p}=\frac{\pi }{4},\mu
_{11}t_{p}=p\pi ,p=1,2,...,$ ($\alpha =4)$ the system is found to be in
maximally entangled tripartite two mode GHZ state \cite{shar103}, 

\begin{equation}
\Psi _{GHZ}=\frac{\left( -1\right) ^{p}}{\sqrt{2}}\left( \left|
g,0,0\right\rangle -i\left| e,1,1\right\rangle \right) .  \label{2.10}
\end{equation}
Using Eq. (\ref{2.9}), we can easily write down the density operator $%
\widehat{\rho }(t)$ for $\widehat{\rho }(0)=\left| g,0,0\right\rangle
\left\langle g,0,0\right| $. Next we evaluate the probability $P_{GHZ}(t)=tr(%
\widehat{\rho }(t)\widehat{\rho }_{GHZ})$ of finding the system in the state 
$\widehat{\rho }_{GHZ}=$ $\left| \Psi _{GHZ}\right\rangle \left\langle \Psi
_{GHZ}\right| $\bigskip\ and obtain 
\begin{widetext}
\begin{eqnarray}
P_{GHZ}(t) &=&\frac{1}{2}.+\frac{\Omega ^{2}}{2\mu _{11}^{2}}\left( \exp
\left( \frac{-2\mu _{11}^{2}t}{\gamma }\right) \cos \left( 2\mu
_{11}t\right) \right)   \nonumber \\
&&+\frac{\Omega ^{2}}{2\mu _{11}^{2}}\left( \exp \left( \frac{-2a_{11}^{2}t}{%
\gamma }\right) \sin \left( 2a_{11}t\right) -1\right)   \nonumber \\
&&+\frac{1}{2}\left( \frac{\mu _{11}-a_{11}}{2\mu _{11}}\right) ^{2}\exp
\left( \frac{-2\left( \mu _{11}+a_{11}\right) t}{\gamma }\right) \sin \left[
2(\mu _{11}+a_{11})t\right]   \nonumber \\
&&-\frac{1}{2}\left( \frac{\mu _{11}+a_{11}}{2\mu _{11}}\right) ^{2}\exp
\left( \frac{-2\left( \mu _{11}-a_{11}\right) t}{\gamma }\right) \sin \left[
2(\mu _{11}-a_{11})t\right] .  \label{2.11}
\end{eqnarray}
\end{widetext}

\begin{table}
\caption{$P_{GHZ}(t)$ peak values at $T=(\pi /4,3\pi /4),$ for $%
a_{11}/\gamma =0.001,0.005,0.01,0.1$.\label{T1}}
\begin{ruledtabular}
\begin{tabular}{|c|c|c|c|}
\hline
$R$ & $\frac{1}{\gamma }$ & $P_{GHZ}$ & $P_{GHZ}$ \\ \hline
($=a_{11}/\gamma $) & $\ ($in nano sec) & $T=\pi /4$ & $T=3\pi /4$ \\ \hline
\multicolumn{2}{|c}{No intrinsic decoherence} & $1.0$ & $1.0$ \\ \hline
$0.001$ & $0.43$ & $0.99$ & $0.94$ \\ \hline
$0.005$ & $2.15$ & $0.94$ & $0.78$ \\ \hline
$0.01$ & $4.32$ & $0.89$ & $0.65$ \\ \hline
$0.1$ & $43.20$ & $0.53$ & $0.37$ \\ \hline
\end{tabular}
\end{ruledtabular}
\end{table}

In the limit $a_{11}/\gamma \rightarrow 0$ corresponding to continuous time
evolution$,$ Eq. (\ref{2.9}) and Eq. (\ref{2.11}) reduce to decoherence free
time evolution of density operator and $P_{GHZ}$ respectively. Fig. 1
displays $P_{GHZ}(t)$ as a function of scaled time variable $T(=a_{11}t)$,
for the choice $\mu _{11}\backslash a_{11}=4,$ and intrinsic decoherence
parameter $R=a_{11}/\gamma =0.001,0.005,0.01,0.1$. Evidently the evolution
dynamics of $P_{GHZ}(t)$ is sensitive to changes in intrinsic decoherence
parameter $R$. With increase in the value of $R$ the peaks in $P_{GHZ}(t)$
plots are seen to get lower. Using the value $\Omega =8.95MHz$ \cite{mund03}
and recalling that $\mu _{mn}=\sqrt{a_{mn}^{2}+\Omega ^{2}}$ , we calculate
the mean frequency $\gamma $ of minimum time step. Table 1. lists the $%
P_{GHZ}(t)$ peak values at $T=\pi /4,3\pi /4,$ for the choices $%
a_{11}/\gamma =0.001,0.005,0.01,0.1.$ As expected, the presence of decay
factors in Eq. (\ref{2.11}) causes the $P_{GHZ}(t)$ peak values to decrease
with increase in $1/\gamma .$ For $a_{11}/\gamma =0.1,$decoherence effects
dominate the scene resulting in a low $P_{GHZ}(t)$ peak value of $0.53$. We
may recall here that for the coupling value determined by $\Omega =8.95MHz,$
the angle $T=\pi /4$ corresponds to an interaction time of $t_{1}\sim
0.34\mu s.$ The probability of finding the system in GHZ state can be
measured by cavity-photon measurement combined with atomic population
inversion measurement.

\section{Conclusions}

We have analysed the evolution dynamics of a single two-level trapped ion
coupled to an optical cavity interacting with a laser field and a single
cavity mode. Assuming that the quantum system evolves in a stochastic
sequence of identical unitary transformations \cite{milb91}, the probability
of generating a maximally entangled tripartite GHZ state has been
calculated. We have used intraction parameters reported in a recent
experiment \cite{mund03}. As expected, the presence of decay factors in time
evolution equation, causes the $P_{GHZ}(t)$ peak values to decrease with
increase in the average size of the minimum time step. The evolution
dynamics of $P_{GHZ}(t)$ is also sensitive to changes in intrinsic
decoherence parameter $R$. Besides that the energy spectrum of system
hamiltonian determines the decoherence rate for a given initial state
preparation. For the case at hand, that is ion initially in its ground state
occupying the lowest energy trap level, while the cavity is prepared in
vacuum state, the ratio $R=(\frac{1}{2}g\eta _{c})/\gamma $ is crucial in
determining the rate at which GHZ state generation probability $P_{GHZ}(t),$
decays with time. Even for a large value of $a_{11}/\gamma =0.01$, we obtain
GHZ state generation probability of $\ 0.89.$ Decoherence effects due to
cavity decay, heating and other random sources of decoherence on $P_{GHZ}$
have not been considered.

\begin{figure}[t]
\centering
\includegraphics[width=2.2in,height=3in,angle=-90]{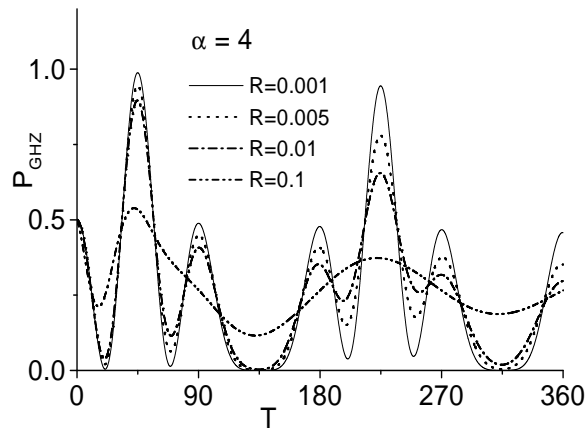}
\caption{$P_{GHZ}(t)$ as a function of scaled time variable $T(=a_{11}t)$ in degrees,
for the choice $\mu _{11}\backslash a_{11}=4,$ and intrinsic decoherence
parameter $R=a_{11}/\gamma =0.001,0.005,0.01,0.1$. \label{fig1}}
\end{figure}

\end{document}